| | Digital Twins and Applications |

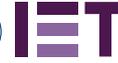 The Institution of Engineering and Technology  WILEY

**PERSPECTIVE**

# On future power system digital twins: A vision towards a standard architecture

Wouter Zomerdijk[1] ⓘ | Peter Palensky[1] ⓘ | Tarek AlSkaif[2] ⓘ | Pedro P. Vergara[1] ⓘ

[1]Intelligent Electrical Power Grids, Delft University of Technology, Delft, The Netherlands

[2]Information Technology Group, Wageningen University and Research, Wageningen, The Netherlands

**Correspondence**

Wouter Zomerdijk, Intelligent Electrical Power Grids, Delft University of Technology, Mekelweg 4, 2628 CD, Delft, The Netherlands.
Email: W.Zomerdijk@tudelft.nl

**Funding information**

Rijksdienst voor Ondernemend Nederland, Grant/Award Number: MOOOI32019

**Abstract**

The energy sector's digital transformation brings mutually dependent communication and energy infrastructure, tightening the relationship between the physical and the digital world. Digital twins (DT) are the key concept for this. This paper initially discusses the evolution of the DT concept across various engineering applications before narrowing its focus to the power systems domain. By reviewing different definitions and applications, the authors present a new definition of DTs specifically tailored to power systems. Based on the proposed definition and extensive deliberations and consultations with distribution system operators, energy traders, and municipalities, the authors introduce a vision of a standard DT ecosystem architecture that offers services beyond real-time updates and can seamlessly integrate with existing transmission and distribution system operators' processes while reconciling with concepts such as microgrids and local energy communities based on a system-of-systems view. The authors also discuss their vision related to the integration of power system DTs into various phases of the system's life cycle, such as long-term planning, emphasising challenges that remain to be addressed, such as managing measurement and model errors, and uncertainty propagation. Finally, the authors present their vision of how artificial intelligence and machine learning can enhance several power systems DT modules established in the proposed architecture.

## 1 | INTRODUCTION

The introduction of the supervisory control and data acquisition (SCADA) system and the wide area monitoring system (WAMS) are perfect examples of how digitalisation in power systems has led to an increase in system reliability over the past decades.[1] Enabled by the digital paradigm shift, investments in information and communication technology (ICT) have enabled power system operators to install advanced metring infrastructure (AMI), perform real-time (partial) network monitoring, and enhance simulation accuracy.[2] Consequently, facilitating more precise predictions of the system response and thus enhancing operational efficiency.[3] However, the increasing size and complexity of power systems, the surge in distributed energy resources (DERs), and the transformation towards flexible and controllable loads require the management of immense data pools, advanced simulation and modelling efforts, and the development of more complex operation and planning approaches. A digital twin (DT) can provide a holistic approach to data processing, modelling, simulation, and service validation,[4] thereby playing an essential role in bridging the gap between physical and digital models.[5]

The DT concept was initially imagined over 3 decades ago in Ref. 6 and adapted to product life cycle management several years later.[7,8] As a concept, NASA matured DTs for space exploration using remote-controlled vehicles.[9] At that time, a DT was defined as "*an integrated multiphysics, multiscale, probabilistic simulation of an as-built vehicle or system that uses the best available physical models, sensor updates, fleet history etc., to mirror the life of its corresponding flying* twin".[10] In this definition, the DT's main features comprise an ultra-high-fidelity physical model, real-time monitoring of health and performance, and data mining. With such features, NASA's DT can self-adapt, forecast future states, predict system responses, and mitigate damages.

Since then, the DT concept has been adapted to the power systems domain to evolve into a twin-centric digital control centre architecture[4,11] in which the physical system and the dynamic simulation-based digital world are interlinked by a real-time automated data flow. The capabilities of such a system include anomaly detection and threat mitigation, self-adaptation, and prediction of the system response, similar to the capabilities first envisioned in Ref. 12. Despite years, the general DT concept in the power systems domain remains vague and surrounded by misconceptions, as exemplified in Figure 1.[13] For instance, digital models and shadows are often portrayed as DTs. A simplified schematic representation of both a digital model and a digital shadow is presented in Figure 1a,b, respectively. These representations lack one of the fundamental features of the DT concept, namely a bidirectional real-time automated data exchange between a physical entity (or system) and its digital counterpart. In a digital model, once created, modifications to the physical system do not affect the digital model. Similarly, in a digital shadow, changes in the physical system influence the digital shadow, but not vice versa.[13] Moreover, the key features of DTs are domain-specific and can have different levels of sophistication. In this context, the need for a customised power system DT definition has already been recognised in the literature.[14] Furthermore, researchers acknowledge the importance of DTs without achieving consensus on a universal approach, underlining the development of a standard ecosystem architecture to advance DT research and implementation.[15]

This paper begins by examining the evolution of the DT concept across various engineering domains before narrowing its focus to the power systems domain. Then, it identifies and highlights key features that we consider fundamental for the future development of DTs in power systems. By reviewing different definitions and capabilities, we propose a new standard definition of DTs specifically tailored to power systems. Building on the proposed definition, a vision of a standard DT ecosystem architecture designed to offer services beyond real-time monitoring, control, and operation is proposed. Later, we elaborate on our vision for DTs in advanced power system life cycle phases (i.e. long-term planning), highlighting challenges that remain, such as managing measurement and model errors, and uncertainty propagation. Finally, we present our vision of how artificial intelligence (AI) and machine learning (ML) can enhance several power systems DT components and modules established in the proposed standard ecosystem architecture. In summary, the main contributions of this paper are as follows:

- *Tailored power systems DT definition*: A new standard definition of DTs in power system is proposed. The proposed DT definition extends beyond NASA's DT definition, represented in Figure 1c, including a long-term planning component, among others, as depicted in Figure 1d. It underscores the importance of integrating advanced data management and high-fidelity models for power system operation and planning, thereby aiming to consolidate DT definitions across domains to establish a standard definition for future DT implementation.

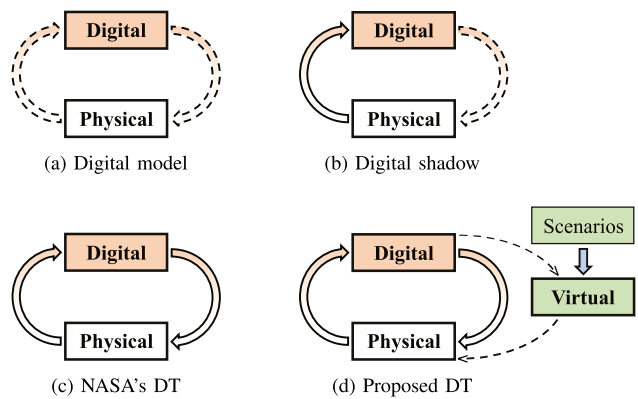

**FIGURE 1** Common misconceptions found in literature, based on a review by authors in Ref. 13. Dashed arrows represent manual data exchange and solid arrows represent real-time automated data exchange between a physical system and its digital counterpart. Digital models and shadows are misconceived as DTs even if they miss such real-time automated data exchange between the physical and digital systems. Expectations of the DT concept go beyond real-time monitoring services.

- *Standard power systems DT architecture*: A standard DT ecosystem architecture is proposed to collect the author's vision on such developments aiming to offer services beyond real-time monitoring, control, and operation. The proposed ecosystem architecture can seamlessly integrate with existing transmission system operator (TSO) and distribution system operator (DSO) processes while reconciling with concepts such as microgrids (MGs) and local energy communities (LECs) based on a system-of-systems view. This architecture resulted from extensive deliberations and consultations with DSOs, energy traders, and municipalities and aims to address their primary enquiries regarding the use of DTs to support their wide range of primary objectives. While empirical validations through case studies are crucial, the current paper serves to set forth a conceptual foundation and roadmap for the development and implementation of the proposed architecture.

- *Addressing power systems DTs as a large-scale software development challenge*: Our proposed envisioned architecture addresses the large-scale software development challenges often overlooked by existing DT platforms, advocating for a standardised approach to manage the complexity and scale of modern power systems. The proposed power system DT ecosystem architecture integrates advanced data management capabilities to address measurement and model errors and uncertainties, enhancing the accuracy and reliability of the DT for robust decision-making support. Moreover, we adopt a system-of-systems perspective, allowing for integrating DTs developed at various levels, ensuring scalability and flexibility in more complex systems.

- *Future-oriented DT vision*: Our proposed DT architecture incorporates AI and ML to enhance data processing, predictive maintenance, operational optimisation etc., offering advanced predictive capabilities and operational efficiencies as envisioned enhanced features of future power systems DTs platforms.

## 2 | DIGITAL TWINS AND POWER SYSTEMS

The prospective capabilities of DTs depicted in the NASA papers[9,10] laid the foundation for extensive research on the application of DTs and their widespread adoption in different industries. Variants of NASA's definition have been proposed for both generic purposes and specific domains, especially in applications for manufacturing and mechatronics. The International Academy for Production Engineering (CIRP) proposes a DT definition for generic purposes that encompasses similar features to NASA's definition. However, these features can be tailored for specific purposes.[16] By delineating a DT as a system with a wide range of tailorable features, it allows for systems ranging from a static human-triggered simulation tool to a dynamic fully autonomous forecasting tool to be labelled as a DT. Therefore, the tailorable DT definition provided by CIRP comes at the expense of the fundamental paradigm of twinning, that is, a bidirectional real-time automated data exchange between a digital and physical entity.[17]

### 2.1 | Digital twin definitions and features

In general, NASA's and CIRP's definitions highlight three key DT features: a *physical entity* (or system), a *digital entity* (or system), and their *data exchange*, as can be seen in Figure 2. Although the CIRP definition recognises the importance of a physical-to-digital connection, that connection need not necessarily be bidirectional, real time, and automated. Recognising these as the main features enables DTs to offer their well-known *real-time monitoring* services. Nevertheless, driven by the increased complexity of engineering systems and the involved ICT infrastructure required to handle large amounts of data, additional features have been gaining more attention. The additional features outside the power systems domain are summarised in the first seven entries of Table 1 and elaborated on in Section 2.1. They include *services and modules* that go beyond real-time monitoring, encompassing tasks such as scenario simulation, system optimisation, diagnostics, and performance predictions.[9,10,18] Indeed, the deployment of these advanced services and modules requires standardised ways of sharing data using *data interfaces* as well as more complex *data storage* and *data visualisation* capabilities. As a result, these data capabilities are now recognised as important DT features,[19] as shown in Figure 2.

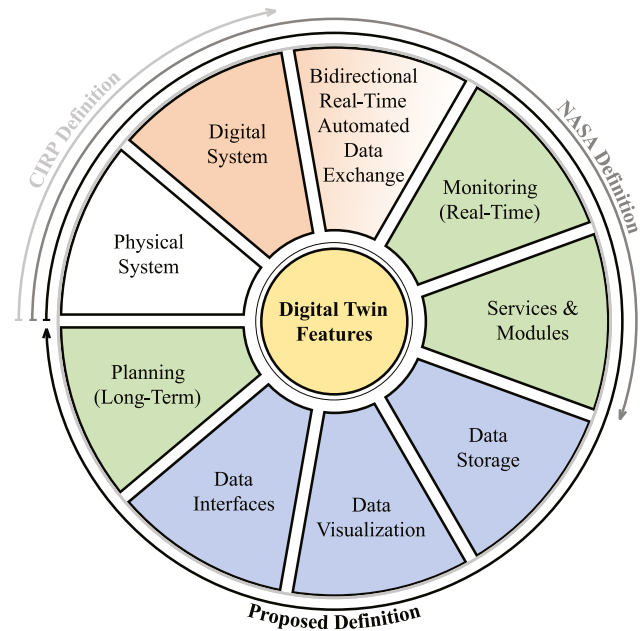

**FIGURE 2** Key features recognised as part of the proposed DT definition. Although the CIRP definition recognises the importance of a physical-to-digital connection, that connection need not necessarily be bidirectional, real time, and automated. The original NASA definition recognised the first three features (a physical system with a digital representation and a bidirectional real-time automated data exchange), enabling real-time monitoring and short-term state prediction. Updated definitions recently recognise the need for new features, such as those supporting more advanced data sharing, storage, and visualisation. We argue for the inclusion of (long-term) planning as a new feature, expanding DT services beyond real-time monitoring and their deployment within the power systems domain. The features in the figure follow a colour coding that will be used throughout this paper.

Despite the comprehensive list of features now recognised as fundamental for a DT, DTs are domain-specific and can have different levels of sophistication. They are generally designed to be embedded into domain-specific processes within a broader system's life cycle (i.e. conceptualisation and design, operation, and planning).[20–22] From an engineering system's perspective, DTs are expected to provide services and support beyond real-time updates and monitoring. For instance, they can be implemented in the design phase and guide planning decisions in the long-term planning phase.[23] Due to the range of requirements in the different engineering system's life cycle phases, we argue that the inclusion of *(long-term) planning* services is crucial in the DT definition, and as such, it must be included in Figure 2. This is supported by authors in Ref. [24] that state that incorporating DTs at advanced life cycle phases (i.e. long-term decision-making) guides the design of the DT itself. Moreover, a mature DT definition facilitates the integration of such an ecosystem into already established operation and planning frameworks of any engineering system. Nevertheless, industrial applications of DTs at advanced life cycle phases remain limited or are in their conceptual phase.

**TABLE 1**   DT key features in available definitions and system architectures.

| Ref. | Physical system | Digital system | Data exchange* | Monitoring (real-time) | Services and modules | Data storage | Data visualisation | Data interfaces | Planning (long-term) | Domain |
|---|---|---|---|---|---|---|---|---|---|---|
| 9,10 | ✓ | ✓ | ✓ | ✓ | ✓ | | | | | Manufacturing |
| 16 | ✓ | ✓ | □ | □ | □ | | | | | Manufacturing |
| 18 | ✓ | ✓ | ✓ | ✓ | ✓ | | | | | Manufacturing |
| 19 | ✓ | ✓ | ✓ | ✓ | ✓ | ✓ | ✓ | ✓ | | Manufacturing |
| 20–22 | ✓ | ✓ | ✓ | ✓ | ✓ | ✓ | | ✓ | | Mechatronics |
| 23 | ✓ | ✓ | ✓ | ✓ | ✓ | ✓ | ✓ | ✓ | | Mechatronics |
| 24 | ✓ | ✓ | □ | □ | □ | | | | | MBSE |
| 25 | ✓ | ✓ | ✓ | ✓ | ✓ | | | | | Power system |
| 4,11 | ✓ | ✓ | ✓ | ✓ | ✓ | | | | | Control centre |
| 26 | ✓ | ✓ | ✓ | ✓ | | | | | | Control centre |
| 27 | ✓ | ✓ | ✓ | ✓ | ✓ | | | | | Control centre |
| 28 | ✓ | ✓ | ✓ | ✓ | | | | | | Component |
| 29 | ✓ | ✓ | ✓ | ✓ | | | | | | Component |
| 30 | ✓ | ✓ | ✓ | ✓ | ✓ | | | | | Component |
| 31 | ✓ | ✓ | ✓ | ✓ | ✓ | | | | | Component |
| 32 | ✓ | ✓ | ✓ | ✓ | ✓ | | ✓ | | | Component |
| 33 | ✓ | ✓ | ✓ | ✓ | ✓ | ✓ | ✓ | ✓ | | Component |
| 34 | ✓ | ✓ | ✓ | ✓ | ✓ | | | | ✓ | Microgrid |

*Note*: * bidirectional, real-time, and automated; ✓, included; □, tailorable feature.

## 2.2 | Power system digital twins

In the power systems domain, DT architectures highlight similar features described in the general DT definition, that is, a physical system, a digital system, and their interconnection.[25] See, for example, the CIRP and NASA definitions in Figure 2. The DT definitions and features found in this domain are summarised in the last eleven entries of Table 1 and elaborated on in Section 2.2. At the TSO level, such DT features are straightforward to identify, with SCADA systems, energy management systems (EMSs), and WAMSs supporting the bidirectional data connection of assets and resources with their digital counterparts,[4,11] enabling real-time monitoring as a base for several modules. At the DSO level, the massive deployment of Internet of things (IoT) sensors and smart metres facilitates data exchange, enabling real-time response for large networks[26] and advanced DER management deployment.[27] At the component level,[28,29] developed a DT for medium voltage (MV) to low voltage (LV) transformers and dc-dc power converters, respectively. By combining high-fidelity simulation models of transformer and converter components with measurement devices, the number of measurement devices needed to support accurate real-time monitoring was minimised. The DT presented in Ref. 30 analyses operational data in order to increase the transformer's lifespan by ensuring safety under future operating conditions. To improve anomaly detection and maintenance scheduling, Ref. 31 develops a DT for a power plant, implementing an architecture comprised of a continuous data stream from sensors and physics-based dynamic models. In a similar power plant application, Ref. 32 develops a DT for real-time monitoring and control. Unlike Refs. 31 and 32, highlights the importance of data visualisation to facilitate monitoring and control by the power plant operator, although only limited visualisation capabilities (i.e. industrial charts) are considered. Exploiting a more software-based architecture, Ref. 33 proposed using DTs as cloud battery management systems (BMSs) to replace the onboard BMS. This improves the computational capabilities of these systems and enables big data storage, advanced visualisation, and reliable system prediction and optimisation. Contrary to the perspective presented in Ref. 33, even though data storage, visualisation, and interfaces are deemed necessary, they are not universally acknowledged as inherent to the DT concept, which underpins the development of the aforementioned features. Moreover, applications of DTs in long-term system planning are scarce.[34]

From the deployment perspective, two issues remain unclear: (1) how DTs can be embedded into the power system's life cycle and (2) how the power system DT definition can reconcile with similar DTs developed on smaller scales (e.g. component level DTs),[35] as well as new concepts such as microgrids (MGs) and local energy communities (LECs).

Considering this, and as most of the current DT applications in the power systems domain overlook several established DT features, a proper power system DT definition is missing. This definition must consider all recognised features in Figure 2 and integrate the specific modules and models used to control, operate, and plan power systems. We propose a new DT definition tailored to power systems to address this limitation.

## 2.3 | A new power system digital twin definition

The typical DT definition, that is, a physical system with a digital representation and a bidirectional real-time automated data exchange, does not fully represent the range of services and modules in the power systems domain. Furthermore, this definition poses challenges in terms of effectively integrating DTs into the established frameworks governing TSOs and DSOs across control (real time, at the seconds horizon), operation (at the minutes, days, and weeks horizon), and planning (at the years horizon) tasks. Although customised power system DT architectures have been developed based on application-oriented definitions (see Table 1), they disregard that (1) the deployment of DTs is mainly a large-scale software development problem that requires a standardised approach, (2) the non-stationary environment inherent to power systems requires self-adapting models (without this representing a challenge), (3) power systems are composed of the interconnection of different assets (e.g. transmission lines and transformers) that can also be modelled via complex and detailed simulation models or DTs, and (4) measurement and model errors and uncertainties propagate through the entire life cycle. Considering such features, we propose a new DT definition customised to power systems and aligned with system operators' perspective. We define a power system DT as *a collection of modules and models (based on multiphysics simulation) integrated into a single software ecosystem with advanced data management and visualisation capabilities aimed at mirroring the real-time operation of the power system and supporting its long-term planning.*

The aforementioned definition underscores the necessity of considering several (software) modules (e.g. voltage control and security assessment) and models (e.g. power flow and state estimation) used in power system operation and control as well as software modules that enable data processing and storage, data generation, forecasting, and data visualisation. Moreover, it incorporates leveraging the DT for tasks that go beyond the operation framework, such as assessing future operational scenarios (e.g. utilising the power flow model and forecasted data) and supporting long-term infrastructure upgrades. Defining a power system DT as a collection of integrated modules within a single (software) ecosystem facilitates the further incorporation of DTs developed at different scales (e.g. at the component level), provided that a proper interface is deployed. To complement the presented definition and facilitate its integration into TSOs' and DSOs' processes, a standardised architecture is still needed. In the next section, we present and discuss our vision of a power system DT ecosystem architecture while highlighting undressing technical challenges to deploy such architecture.

## 3 | ENVISIONED STANDARD POWER SYSTEM DIGITAL TWIN ECOSYSTEM ARCHITECTURE

We conceive the implementation of power system DTs as a comprehensive software ecosystem development and deployment challenge, necessitating a holistic approach. As a result, a standardised ecosystem architecture is needed that acknowledges the DT's main features already discussed and presented in Figure 2 while enabling seamless integration into one software ecosystem. The envisioned and proposed DT ecosystem architecture, shown in Figure 3, is developed from the operators' perspective and aligned with their operation frameworks. Nevertheless, it is general enough to be easily extended to other applications. The envisioned architecture in Figure 3 is composed of five main components: (1) *High-Fidelity Simulation Models*, (2) *Operation and Planning*, (3) *Grid as a Service*, (4) *Data Engineering*, and (5) *Data Analytics*. The former components aim for an autonomous power system operation, enabling operators to meet their objectives from everyday system operation to long-term planning, as well as supporting external businesses and stakeholders, while the latter components are the backbone, allowing bidirectional real-time automated data exchange between the digital and physical world, as well as between the different components, modules and models via proper data interfaces. The envisioned ecosystem architecture interfaces (with appropriate data interfaces) with the TSOs' and DSOs' control centres. In this view, we do not foresee that such a power system DT will replace the operators' centres. Nevertheless, we envision it to evolve to be a crucial support tool.

## 3.1 | High-fidelity simulation models

Multiphysics, multiscale, probabilistic simulation models compose the heart of DTs and support all the capabilities of the different DT modules. From the power systems' perspective, the *High-Fidelity Simulation Models* component comprises simulation models such as *Power Flow Models* for steady-state analysis, *EMT Models* for electromagnetic transients (EMT), *Thermal Models* for heat analysis, and *Dynamic Phasor Models* for analysis based on phasor measurement units (PMUs).[37] These models are used to increase the system operators' observability and control capabilities of their networks. Advanced simulation models can also be incorporated, such as *Hardware-in-the-Loop (HiL) Models* using real-time digital simulators. The literature on power system modelling and simulation tools is vast,[38] and detailed taxonomies per model application are already available.[39] Due to the non-stationary environment in which a power system operates, a continuous update process is expected to be in place. This can be implemented using the *Simulation Models Management* module, in charge of keeping track of infrastructure (e.g. new assets) or operation changes (e.g., network topology change). Moreover, it ensures time synchronisation among the different simulation models, enabling them to cooperate and simulate using data with varying time resolutions. Proper model version documentation must also be deployed with the support of the *Data Engineering* component, as detailed in Section 3.3.

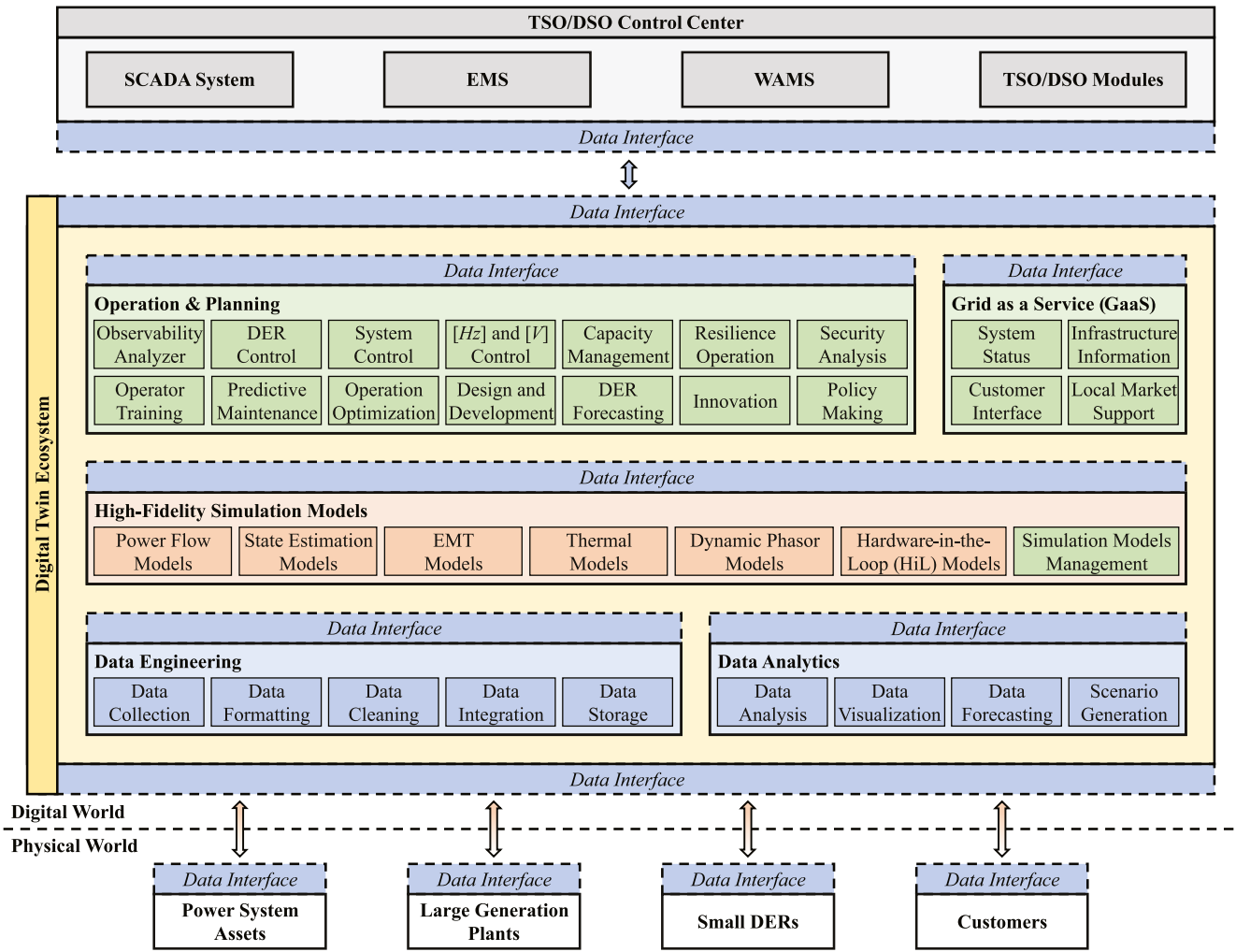

**FIGURE 3** Proposed DT ecosystem architecture based on the proposed DT definition in Section 2. This ecosystem architecture is developed from the operators' perspective and aligned with their operation frameworks. Nevertheless, it is general enough to be easily extended to other applications. The white blocks represent the most relevant physical systems, blue blocks represent data modules, orange blocks represent high-fidelity simulation models, green blocks represent some of the power system operation and planning modules, as described in,[36] grey blocks represent some of the systems and processes at TSOs' and DSOs' control centres, and arrows represent data (and information) flows. The components in the figure follow the same colour coding throughout the paper.

## 3.2 | Bidirectional real-time automated data exchange

Digital infrastructure serves as the foundation for the digital-to-physical connection. To be fed into the DT, data is collected by field measurements, IoT devices, and smart metres from several power system assets (e.g. lines, switches, and transformers), large generation plants (e.g. nuclear plants and hydroelectric plans), customers, and DERs (e.g. solar panels, batteries, and electric vehicles). Furthermore, relevant information (e.g. meteorological data, energy market prices, and states of adjacent interconnected systems) composes a set of contextual data and is also collected and stored. In Figure 3, although a direct data connection and exchange is depicted by the DT linking to the power system assets, such data can in practice be transferred via the proper data interface from the TSOs' and DSOs' WAMSs and SCADA systems, located at the operators' premises. In addition to measured data, information in the form of control commands and maintenance recommendations can also flow from the DT to power system assets, generation plants, customers, and DERs. However, handling large volumes of data, including structured, unstructured, and semi-structured data from various sources with different spatio-temporal resolutions, formats, and qualities, poses a challenge.[34] Proper data management is the core function of the *Data Engineering* and *Data Analytics* components.

## 3.3 | Data engineering

The *Data Engineering* component adds to the data processing capabilities of the DT and enables it to collect, clean, integrate, and store data from different sources previously described (see Section 3.2), thereby ensuring that all data are well prepared, accessible, and of high quality for all available modules within the other components. The main modules of this component are *Data Collection*, *Data Formatting*, *Data Cleaning*, *Data Integration*, and *Data Storage*. Typically, power system operators maintain databases with detailed infrastructure information, such as the parameters and locations of power system assets, switching states, and network topology. The *Data Collection* module consists of data systems and interfaces that allow the DT to communicate with these databases and measurement devices to retrieve historical and real-time data on power system assets. A critical aspect is the inclusion of spatio-temporal data, which encompasses time-synchronised field measurements from devices such as PMUs, smart metres, and other measurement devices originating from the WAMSs and SCADA systems. Time-synchronisation of spatio-temporal data is vital for producing reliable simulation results. When combined with geographic information system (GIS) data from TSOs' and DSOs' databases, the data integration module accurately maps all data to the corresponding power system asset within the DT ecosystem. The general term data interface is used as it refers to the point of interaction or communication between different systems, components, modules, and models. The definition of the proper data interface is application- and use-case-specific, but it can include, for instance, a definition of application programming interfaces (APIs), as well as security requirements (e.g. authentication and encryption). The *Data Formatting* module is in charge of properly formatting the data (e.g. JSON, XML, CSV, binary, etc.) to support all DT modules. For instance, the common information model (CIM) is a widely used standard in power systems for defining a common format to describe assets like generators, transformers, and substations.[40,41] Operators adhere to CIM standards for consistent asset naming, ensuring interoperability across systems and modules. The *Data Cleaning* module checks the veracity of the data and complements, corrects, or discards data where needed, ensuring high data quality. For instance, data cleaning algorithms may detect and remove anomalies in power measurements, such as spikes or missing values, to improve the reliability of analysis and decision-making. Moreover, in instances where sensors are malfunctioning or absent, the module has the capability to identify and estimate these values using historical measurement data and proximate measurements to support simulations. The enhancements that AI and ML can provide in these scenarios are detailed in Section 5. Power systems comprise various assets, sensors, and control systems, producing vast amounts of heterogeneous data. The *Data Integration* module cohesively consolidates data from diverse sources like SCADA systems, smart metres, and weather sensors. This comprehensive integration enables the DT to analyse and optimise power systems holistically. Finally, the *Data Storage* module enables the large volume of raw and processed data to be stored efficiently for further use in all DT modules. Operators typically employ databases, data warehouses, and data lakes, including cloud-based storage.

## 3.4 | Data analytics

The *Data Analytics* component involves the exploration, interpretation, and visualisation of data to derive meaningful insights from the power system operators. In general, data analytics operates across four levels, namely descriptive, diagnostic, predictive, and prescriptive. The first level describes what happened in the past regarding a certain event or phenomenon. The second level diagnoses why this happened. The third and fourth levels, respectively, predict what will happen and optimise operation based on the prediction. Within the proposed ecosystem architecture, the first two levels of data analytics are covered in the *Data Analysis* and *Data Visualisation* modules. These modules analyse processed data from the data engineering component and output data from the operation and planning component. For instance, detecting anomalies, identifying system states, evaluating control actions, and visualising performance indicators are among the capabilities of the underlying models. This enables the DT to describe and diagnose the entire system for its operators. The third level of data analytics is handled by the *Data Forecasting* and *Scenario Generation* modules, which predict future system states on different timescales. The former operates from multiple days ahead to the intraday timescale, while the latter focuses on months to years ahead the timescale. Leveraging historical, meteorological, and socio-economic data, these applications employ statistical, ML, and heuristic models for forecasting. The fourth level is covered by the modules in the operation and planning component. It enables predictive maintenance and optimised decision-making, enhancing system reliability, efficiency, and resilience.[42] Additionally, predictive and descriptive analytics facilitate trend analysis, load forecasting and risk assessment, empowering operators to proactively manage grid operations and address emerging challenges.[43]

## 3.5 | Operation and planning

The *Operation and Planning* component gathers all the already established modules run and executed by the TSOs and DSOs. Modules needed for control and operation, such as *DER Control*, *Observability Analyser*, *System Control*, *Capacity Management*, as well as modules needed for long-term planning, such as *Security Analysis*, *Predictive Maintenance*, *Resilience Operation*, are all deployed into this component, retrieving and exploiting the required data to optimally function via the *Data Engineering* and *Data Analytics* components, respectively. A detailed discussion of each of these well-known and established power system modules is outside the scope of this paper. Nevertheless, interested readers can find a comprehensive view of these modules offered by DSOs in Ref. 36. Notice that in Figure 3, the TSOs' and DSOs' modules are located at their premises. We envision system operators copying such operation and planning modules into the power system DT ecosystem to develop and validate new services by taking advantage of the DT's high-fidelity models. Once validated, they can be deployed back at the system operator control centre premises. In this sense, the power system DT can become the heart of new developments. Nevertheless, this differs from the perspective of some operators,[44] who expect the DTs to become the heart of their control centres.

## 3.6 | Grid as a service

The *Grid as a Service* (*GaaS*) component can be seen as the operators' interface to the external world. In the energy transition context, communities are organising themselves as LECs,[45] while operators of more complex energy infrastructures (e.g. large industrial, military, or university campuses) are constituting MGs. Although LECs and MGs are entirely independent of the operators from the administrative point of view, they are still connected to the power system infrastructure. Moreover, since LECs and MGs can provide crucial support services (e.g. frequency regulation and demand response[46]), cooperation with them is of interest to system operators. Through deployment within the GaaS component, modules such as *Local Markets Support* enable system operators to provide updated information to LECs, thereby promoting smooth local market operations. This could facilitate, for instance, that assets' nominal capacities are respected during energy trading (e.g. via peer-to-peer services[47]) between two independent customers connected in a DSO's operated network. Similarly, modules such as *Customer Interface* can be used to coordinate operations within and between MGs, large customers, aggregators, and charging point operators, supporting the development of new business models. Such coordination can be in the form of validation of their energy schedule, aiming to avoid technical issues in the hosting grid (e.g. congestion). In the case of communicating and sharing updated information with the municipal and regional governments, real-time data may be of less interest. Nevertheless, modules such as *System Status* and *Infrastructure Information* can be deployed to share global system information in terms of pre-defined performance indices (e.g. minimum, average, and maximum utilisation rate of transformers). Such governments can use this information to gain better insights into the city's electricity infrastructure and develop customised urban plans faster.

From the system operators' perspective, integrating a dedicated component into the DT for communication and data exchange with external entities, as previously discussed, enhances coordination while safeguarding the integrity of the TSOs' and DSOs' platforms. Maintaining platform integrity is crucial when interfacing with external parties as it mitigates the risk of cyberattacks on the TSOs and DSOs.[48] Moreover, as a separate component, it offers enough flexibility to the system operators to build new and customised modules on demand.

For instance, a new capacity market module can be developed to interface the DSO with customers' trading capacity as a commodity. Such markets are still in their conceptual phase,[49] and it is unclear how they would be embedded into existing markets.

## 3.7 | System of systems: Multi-digital-twin models

DSOs are responsible for smaller sections of a more extensive network that interfaces with the national transmission network. At the same time, TSOs are responsible for national networks that may go beyond national borders (e.g. electricity interconnection in Europe). Naturally, DTs can be developed for all this infrastructure at different levels by tailoring the DT ecosystem architecture presented in Figure 3. In this context, we envision a natural fit to the system of the systems concept.[50] The concept was adapted to the power systems domain by Ref. 35 in which DTs of lower-level infrastructure, such as lines and transformers, are interfaced with DTs developed for higher levels, such as DTs of regional distribution networks. A simple example of this concept is presented in Figure 4. From the DT ecosystem architecture point of view (see Figure 3), the interface of DTs at lower and higher levels should occur at the *High-Fidelity Simulation Models* component by replacing the corresponding lower level model by a tailored interface to its DT. Such a tailored interface must share the appropriate data depending on the related DT modules and the embedded power systems' life cycle phase (see Section 4 for a more detailed discussion). In this sense, the coupling of multiple DTs via the proposed ecosystem architecture goes beyond the co-simulation concepts already presented for some DT applications,[51] as models are kept intrinsic to each DT. Nevertheless, to achieve this, data standardisation, interfaces, and data exchange efforts are fundamental and should simplify and pave the way for DT implementation. A notable example in this regard is the utilisation of functional mockup units (FMUs), where simulation models are packaged according to a standardised approach, encapsulating the model's equations and its corresponding solver,[52] enabling simulations using different software platforms. Another example of data standardisation and interfaces is the open platform of Vorto,[53] enabling the integration of IoT devices in DTs. However, to achieve large-scale power system DT implementation requires mature system engineering design methodologies. These can be delivered by working together with the software engineering and computer science communities.

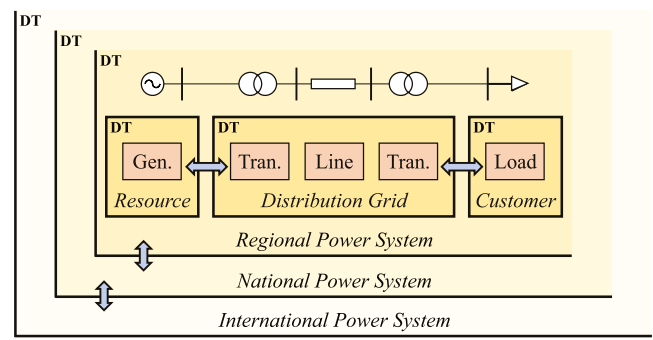

**FIGURE 4** Example of a (small-scale) power system DT frame using the system of systems concept. The orange blocks represent simulation models of the physical assets and resources presented by their corresponding white symbols in the network diagram. The simulation models are enclosed in DTs represented by yellow boxes (for simplicity, other DT components as presented in Figure 3 are omitted), while the blue arrows represent a bidirectional data exchange between the different levels DTs. The components in the figure follow the same colour coding throughout the paper.

# 4 | LIFE CYCLE OF POWER SYSTEM DIGITAL TWINS

The proposed ecosystem architecture in Section 3 reconciles the key features of the proposed DT definition presented in Figure 2 in the power systems domain while highlighting its software components as the backbone of such a model. Nevertheless, this ecosystem architecture does not depict how to embed such DTs into the power system life cycle, thus overlooking the temporal dimensions. Unlike other systems with clear end-of-life or end-of-cycle phases, the life cycle of a power system does not have a definitive endpoint. As a fundamental infrastructure that supports society, the long-term goal of a power system is to ensure, with high reliability and resiliency, the supply of energy to end customers. In this context, the original product life cycle management view for which DTs were initially proposed[7,8] does not fit with the continuous upgrade and expansion of the power systems' infrastructure. As a result, existing DT definitions fail to consider the additional advantages that DTs can offer to the long-term planning of the power system.

The relevant life cycle of a power system DT comprises a continuous cycle of control, operation, and short- and long-term planning processes performed by the system operators. These processes range from control commands for frequency regulation on a horizon of seconds to capacity management on a horizon of hours, from maintenance recommendations on a horizon of months to policy-making and investment planning on a horizon of decades. Due to the different time resolutions involved, and considering the DT as an integrated ecosystem, it is not clear (1) how the modules (defined in the *Operation and Planning* component) should interact and (2) how data exchange (managed by the *Data Engineering* component) between models (available at the *High-Fidelity Simulation Models* component) should occur. We elaborate on these challenges in the following section.

## 4.1 | Life cycle aligned with TSOs' and DSOs' processes

Figure 5 attempts to reconcile the time dimension of power system DTs with the DSOs needs of integrating multiple models used for different operation and planning tasks. By highlighting data (and information) flows between models used for tasks located at different time horizons (e.g. seconds, minutes, years, and decades), the envisioned DT architecture and its key components are linked to the continuous cycle of processes performed by the system operators. Nevertheless, integrating data with different time resolutions within these processes and their underlying simulation models poses significant challenges. For example, consider the Dutch household's demand profile with several time resolutions presented in Figure 6. Using demand profiles with different time resolutions as input for the DT simulation models can alter their output due to the inherent information loss, especially considering the stochastic behaviour loss in lower time resolution data. As a result, important dynamics, such as peak demands, can be overlooked, impacting the power system's operational resilience, for instance, by underestimating the peak consumption. Therefore, selecting the appropriate time resolution for each simulation model, performing data downsampling or upsampling as needed, and ensuring time synchronisation of measurement data and simulation models are essential for accurate results. A multi-domain novel generation modelling (NGM) platform proposed in Ref. 54 supports multi-time-scale operation within the same system. Nevertheless, this research area is usually overlooked in the power system community.

As data flows from modules used in control to long-term planning, as depicted in Figure 5, it allows for a data downsampling (or data compression) process to be implemented. For instance, the *EMT Models* (which run at the second timescale) could be initialised by making use of the *Power Flow Models* (which run at the 5- or 15-min timescales). Similarly, the *Security Analysis* module (which runs at the hour timescale to assess operational feasibility) could utilise the *Power Flow Models*. For infrastructure upgrade and expansion, the *Capacity Management* module (which runs at the hour timescale) could utilise a lower time resolution assessment. For instance, by making use of the *Power Flow Models* and *Thermal Models* and building on top of modules such as *DER Control* and *Observability Analyser*. Integrating and properly interfacing all these models and modules within the same DT ecosystem will bring added benefits to system operators. As a result, investment decisions regarding infrastructure upgrades can be assessed at the second timescale within the same ecosystem with minimal technical effort. With these characteristics, we foresee a shift away from the traditional paradigm of segregated operation and planning tasks, enabling decisions made at any stage within the power system's operational cycle to be promptly evaluated across different time resolutions.[55]

## 4.2 | Life cycle challenges

The flow of data between components, modules, and models displayed in Figure 5 poses the risk of facilitating error and uncertainty propagation throughout the entire power system life cycle. Due to the intrinsic complexity of power system DT architectures, these are usually disregarded. Nevertheless, they can largely affect decision-making at different power system's life cycle phases. In the power systems domain, uncertainty is usually considered from the stochastic behaviour of the demand and renewable-based generation, enabling the development of complex stochastic and robust mathematical formulations.[56] Although such uncertainty is relevant, in the context of a power system DT, errors and uncertainties from measurements, data up/downsampling, and errors and inaccuracy from models must also be considered. In this sense, efforts to reduce the gap between simulation and reality have already been recognised in other domains, such as autonomous driving.[57] Nevertheless, efforts in this direction in power systems are not widely known. Considering uncertainty propagation among models with different time resolutions increases the complexity of power system DT developments, raising several important modelling questions yet to be answered. Certainly, this is a future research direction that requires much attention.

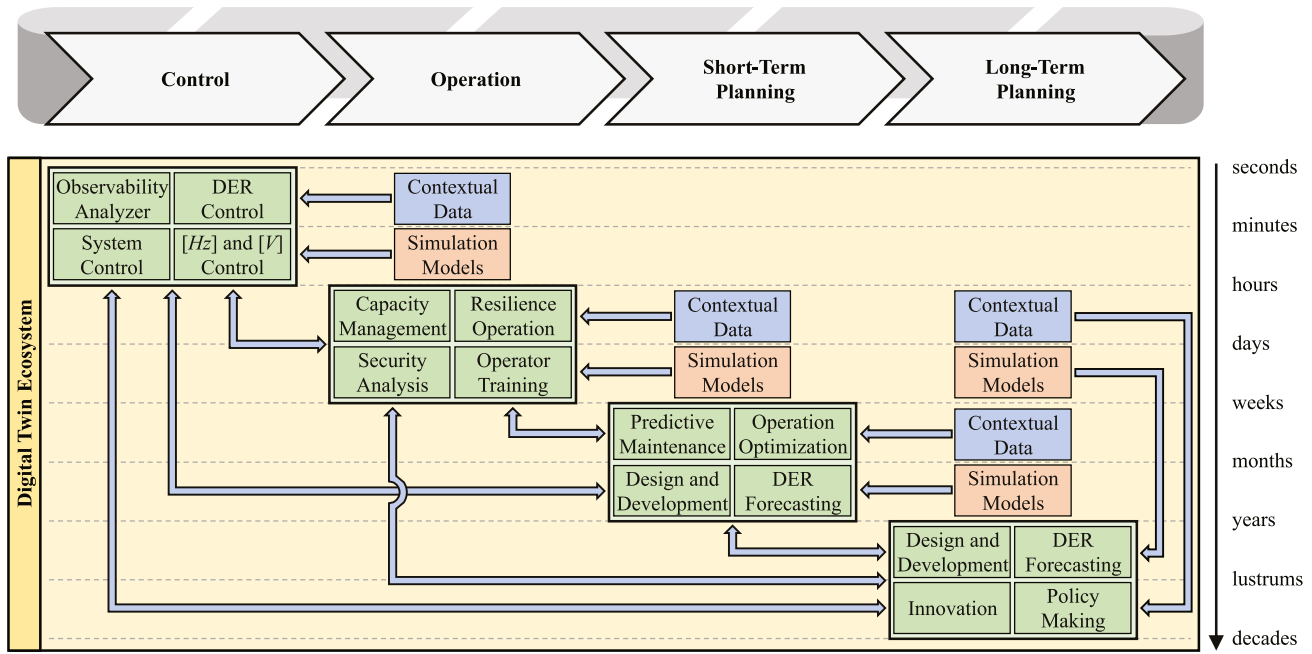

**FIGURE 5** Life cycle of the DT modules related to the continuous cycle of processes performed by system operators. The grey blocks on the top represent the relevant power system life cycle phases and the *y*-axis on the right shows the timescale of the relevant processes. The blue lines/blocks represent data-related processes and modules, the green text blocks represent some of the power system operation and planning modules, while the orange blocks represent high-fidelity simulation models needed for these modules. The components in the figure follow the same colour coding throughout the paper.

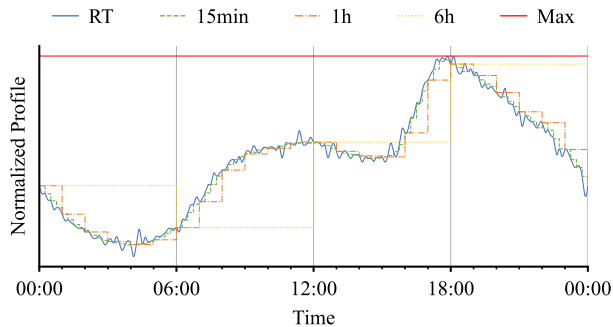

**FIGURE 6** A normalised load profile of a Dutch household presented for several time resolutions. As can be seen, the inherent stochastic behaviour of the load demand is diminished when using a lower time resolution representation. As a result, important dynamics, such as peak demands, can be overlooked. Overlooking important dynamics can impact the power system's operational resilience, for instance, by underestimating the peak consumption.

## 5 | THE ROLE OF ARTIFICIAL INTELLIGENCE AND MACHINE LEARNING IN FUTURE POWER SYSTEM DIGITAL TWINS

In the context of AI, ML models have been proven to successfully learn from large datasets, finding hidden patterns and enabling the discovery of new knowledge. As a result, AI and ML are revolutionising different sectors, from drug production[58] to algorithms development,[59] showing even above human performance in complex games.[60] Similar disruptions are expected in the energy sector, with AI and ML models potentially enhancing the performance of several modules that are the backbone for power system operation and planning.[61] Although several of these advancements have been already recognised,[62] they have not yet been framed to a specific DT architecture. Below, we present our vision (non-comprehensive due to space limitation) of how state-of-the-art AI and ML models can enhance some of DT components and modules described in Section 3. The expected enhancements of specific algorithms, methods, and learning paradigms are divided based on the relevant system operator's processes and are summarised in Table 2.

- *Data Processing and Visualisation*: Power system DTs are expected to collect, store, and process large amounts of data from different energy resources and assets and in different time samples (e.g. seconds, minutes, hours), as well as contextual data (e.g. energy price, weather data). The challenge is properly displaying and processing such a large amount of data, enabling system operators to draw conclusions and insights easily and enabling service provision for external businesses and stakeholders. Processing and visualisation of high-dimensional data are already part of the functionalities within the *Data Engineering* and *Data Analytics* models, and ML models can facilitate them. For instance, natural language processing (NLP) models can be used to analyse and extract insights from large volumes of text data (coming from reports and text file formats) generated by system operators, technicians etc. Furthermore, they can also help standardise and normalise text inputs and even generate summaries, insights, and visualisations from textual data. Other examples are clustering algorithms that can group customers by energy consumption behaviour,[63] while system states can be grouped into representative or typical operation modes.[64] Later, these representative groups and states can be used during operation and planning. Certainly, the data collected and stored by the DT ecosystem will have errors or be incomplete. ML models can automatically identify such data, for instance, using anomaly detection algorithms,[65] while incomplete data can be filled in using ML-based imputation algorithms,[66] such as principal component analysis imputation or deep learning imputation models. All these data correction processes can be automated and executed by the DT ecosystem in the background using these ML models.

- *Modelling*: Multiphysics models use detailed mathematical formulations to simulate the DT's physical system, as discussed in Section 3.1. These complex mathematical models are difficult to solve and may require large computational resources. Therefore, several of the power system DT modules and models (e.g. power flow, optimal power flow, state estimation) will benefit from simplified and complexity-reduced models. ML models have proven to be capable of accurately representing physical systems. For instance, deep learning-based models can be used to solve fast power flow formulations,[67] while graph neural networks (GNNs) exploit the natural graph structure of the power system to accelerate state estimation calculations.[68] The expectation is for deep learning-based power flow solvers to outperform classical formulations, which currently show poor scalability features regarding the network size.[76] In the context of a city-level distribution network DT, many power flow models would need to be scripted. Although modern power flow scripting packages (e.g. PandaPower,[77] Power-GridModel[78]) already standardise input–output formats, and due to the large number of MV and LV networks, their modelling can be a tedious and time-consuming task. Large language models (LLM) can accelerate these scripting tasks by exploiting their automatic programming capabilities.[69] For instance, by deploying an LLM within the DT ecosystem, it can be instructed to learn power flow scripting and automatically provide the large number of power flow models of all the MV and LV networks in a city. Although LLMs are already being proven to automate and support several tasks in various domains,[79] mature applications in the power systems domain remain unseen.[70]

**TABLE 2** Some of the expected enhancements that machine learning models can bring to power system DTs.

| Component | Algorithms/Methods/Learning paradigm | Expected enhancements | Ref. |
|---|---|---|---|
| Data processing and visualisation | NLP | Data processing of large volumes of data | 63 |
| | Clustering | Data visualisation of high-dimensional data | 63 |
| | Clustering and decision tree | Data analysis of operation modes | 64 |
| | Deep learning | Automated and improved anomaly detection | 65 |
| | PCA imputation and deep learning imputation | Imputation of incomplete or bad data | 66 |
| Modelling | Deep learning | Fast power flow formulations | 67 |
| | Deep learning and GNN | Fast state estimation | 68 |
| | LLM | Automated scripting of large and complex power flow models | 69 |
| | LLM | Fast fault and incident reporting | 70 |
| Processes automation | Deep learning | Comprehensive overview of post-switch states | – |
| Operation | RL | Decision-making support for system operators | 71 |
| | RL | Fast control and dispatch optimisation of DERs | 72 |
| Planning | GAN and VAE | Data generation of unseen power system states close to faults | 73 |
| | GAN and VAE | Data augmentation of long-term demand and renewable generation | 74,75 |

- *Processes Automation*: Manually executed processes, such as network switching and work management,[80] can benefit from the fast calculation of the above-mentioned complexity-reduced power flow analysis. Within the *Operation and Planning* component, a new DT module can be developed to run hundreds of network reconfiguration scenarios in the background, providing a good overview of the post-switch state of the system. The system operators can then define the best switching action using this overview. Repetitive, tedious, and time-consuming tasks that require reporting, such as faults and incidents reporting, can be accelerated by using LLMs. LLMs can easily be trained and instructed to generate such reports, reading from the operator's database system information. LLMs can also process current written manuals, providing suggestions for improvement and clarification.

- *Operation*: Voltage optimisation and control, as well as the dispatch of different resources (e.g. batteries), generally rely on the use of optimisation models based on mathematical programming.[81,82] Such optimisation models guarantee mathematical convergence and global optimality (if convexity is proven). However, they suffer from poor scalability and significant computational time.[83] Reinforcement learning (RL) models (the type of ML model used for decision-making) are proven to be capable of solving complex control problems learning either by interacting with the system itself or by using historic decisions.[71,72] RL models are currently being researched in various applications in the power systems domain,[84] and although they do not guarantee global optimality, their generalisation capabilities enable them to provide good quality control actions quickly (once trained). This feature could improve modules deployed within the *Operation and Planning* component, as such RL models can be trained in the background for various decision-making problems, taking advantage of the available data within DTs. In this sense, one of the main features of such RL models is their capability to learn the systems' stochastic behaviour from the data itself, reducing the need to deploy complex stochastic or robust decision-making models, among others.

- *Planning*: Making use of models specialised in data augmentation and data generation, such as generative adversarial networks (GANs) and variational autoencoders (VAEs)[85], rarely seen power system states can be inferred,[73] for example, states close to faults. Such generated representative states can then be used to plan, test, and validate contingency steps, enhancing DTs' fault identification and clearance capabilities. Similarly, data augmentation models can also be used to generate long-term demand and renewable generation time-series data, for example, customer's smart metre readings,[74] as well as market prices.[75] The main advantage of deep learning-based models such as GANs and VAE is that they do not make a priori assumptions of the data probabilistic distribution (e.g. assuming Gaussian distribution). Ultimately, the DTs can use all the synthetically generated time-series data to redefine future planning investments, identifying, for instance, needs in infrastructure upgrades.[86]

Despite the elevated anticipation surrounding AI and ML models for DTs, several challenges still need to be addressed. Large deep-learning models require a large amount of data during training, requiring significant training times and often failing to converge. Depending on the module, such data may not be available or data ownership can be a limitation. Moreover, deep learning models' transparency, trust, and explainability must be enhanced before they can be fully deployed.[87] In this context, we do not envision ML-based models replacing system operators. On the contrary, they will improve the operators' performance, but to achieve this, they must evolve considering system operators in the loop. For instance, using a co-pilot approach as used in the aviation industry.[5]

# 6 | CONCLUSION

This paper proposes a standard definition and ecosystem architecture for power system DTs. The proposed definition reconciles decades of research on the concept and aligns it with the need for further deployment of DTs in the complex domain of power system operation and planning. We argue that a standardised approach allows system operators and software developers to collaboratively develop the backbone of the envisioned DT, that is, a software ecosystem with advanced data management capabilities. The ecosystem's backbone facilitates bidirectional real-time automated data exchange; ensures data standardisation among components modules and models; and supports self-adaptation of all simulation models, among others. The presented power system DT ecosystem architecture enables operators to meet their objectives from everyday system operation to long-term planning, as well as the development and validation of new services by taking advantage of the DT's high-fidelity models. The paper discusses the benefits of connecting all operation and planning modules within a DT ecosystem, as well as multi-DT systems. We propose that the former facilitates the system operators to account for the propagation of errors and uncertainties between these models, and the latter enables more accurate evaluation and long-term infrastructure planning for combined systems. Finally, we anticipate that AI and ML models will play an important role in addressing key bottlenecks in DT development and implementation. These include visualising and processing large amounts of data, accurate and fast modelling of complex systems, and refining future scenario generation. The successful transition of DTs from buzzword to large-scale implementation in power systems depends on the collaboration between system operators and software engineers. We consider that the first step must be taken in the direction of standardisation of workflows and data sharing.

## AUTHOR CONTRIBUTIONS

**Wouter Zomerdijk**: Writing—original draft; writing—review & editing; conceptualisation; investigation; methodology; visualisation. **Peter Palensky**: Supervision; funding acquisition. **Tarek AlSkaif**: Writing—review & editing. **Pedro P. Vergara**: Writing—review & editing; project administration; supervision; funding acquisition.

## ACKNOWLEDGEMENTS

This publication has been funded by the Local Inclusive Future Energy (LIFE) City Project (MOOOI32019), funded by the Ministry of Economic Affairs and Climate and by the Ministry of the Interior and Kingdom Relations of the Netherlands.

## CONFLICT OF INTEREST STATEMENT

The authors declare no conflicts of interest.

## DATA AVAILABILITY STATEMENT

Data sharing not applicable to this article as no datasets were generated or analysed during the current study.

## ORCID

*Wouter Zomerdijk*  https://orcid.org/0000-0003-3463-8363
*Peter Palensky*  https://orcid.org/0000-0003-3183-4705
*Tarek AlSkaif*  https://orcid.org/0000-0002-1780-4553
*Pedro P. Vergara*  https://orcid.org/0000-0003-0852-0169